\documentclass[iop,njp,reprint,superscriptaddress,showkeys,article-title]{revtex4-1} 
\usepackage[utf8]{inputenc} 

\usepackage{cmap} 
\usepackage[T1]{fontenc} 
\usepackage{lmodern}

\usepackage{amsmath,amssymb,mleftright}
\usepackage[version=3]{mhchem} 
\usepackage{graphicx}
\usepackage{bm}
\usepackage{subfigure}

\usepackage{url}
\usepackage{hyperref}
\usepackage[table,usenames,dvipsnames]{xcolor}
\hypersetup{colorlinks=true, linkcolor=BrickRed,
  urlcolor=blue!50!black, citecolor=blue!50!black}

\usepackage{cleveref}
\crefrangelabelformat{equation}{(#3#1#4)$-$(#5#2#6)}

\usepackage[separate-uncertainty=true, multi-part-units=single]{siunitx}
\usepackage{booktabs,tabularx,dcolumn}
\newcolumntype{d}[1]{D{.}{.}{#1}}

\renewcommand\rho{\varrho}

\renewcommand\vec[1]{\boldsymbol{#1}}

\newcommand\diff{\mathrm{d}}
\newcommand\e{\text{e}}

\newcommand\kB{{k_\text{B}}}

\makeatletter
\DeclareRobustCommand*\textsubscript[1]{%
  \@textsubscript{\selectfont#1}}
\def\@textsubscript#1{%
  {\m@th\ensuremath{_{\mbox{\fontsize\sf@size\z@#1}}}}}
\makeatother

\usepackage[normalem]{ulem}

\begin{document}

\title{Active colloidal propulsion over a crystalline surface}

\author{Udit Choudhury}
\thanks{These two authors contributed equally.}
\affiliation{
 Max Planck Institute for Intelligent Systems, Heisenbergstr. 3, 70569 Stuttgart, Germany
}
\affiliation{University of Groningen, Nijenborgh 4, 9747 AG Groningen, The Netherlands}

\author{Arthur V. Straube}
\thanks{These two authors contributed equally.}
\affiliation{
 Department of Mathematics and Computer Science, Freie Universit{\"a}t Berlin, Arnimallee 6, 14195 Berlin, Germany
}

\author{Peer Fischer}
\affiliation{
 Max Planck Institute for Intelligent Systems, Heisenbergstr. 3, 70569 Stuttgart, Germany
}
\affiliation{
 Institute for Physical Chemistry, University of Stuttgart, Pfaffenwaldring 55, 70569 Stuttgart, Germany
}

\author{John G. Gibbs}
\email{john.gibbs@nau.edu}
\affiliation{
 Max Planck Institute for Intelligent Systems, Heisenbergstr. 3, 70569 Stuttgart, Germany
}
\affiliation
{Department of Physics and Astronomy, Northern Arizona University, Flagstaff, AZ 86011, USA}

\author{Felix H{\"o}f{}ling}
\email{f.hoefling@fu-berlin.de}
\affiliation{
 Max Planck Institute for Intelligent Systems, Heisenbergstr. 3, 70569 Stuttgart, Germany
}
\affiliation{
 Department of Mathematics and Computer Science, Freie Universit{\"a}t Berlin, Arnimallee 6, 14195 Berlin, Germany
}
\affiliation{
 Institute for Theoretical Physics IV, University of Stuttgart, Pfaffenwaldring 57, 70569 Stuttgart, Germany
}

\date{\today}

\begin{abstract}
We study both experimentally and theoretically the dynamics of chemically
self-propelled Janus colloids moving atop a two-dimensional crystalline
surface. The surface is a hexagonally close-packed monolayer of colloidal
particles of the same size as the mobile one.  The dynamics of the
self-propelled colloid reflects the competition between hindered diffusion due
to the periodic surface and enhanced diffusion due to active motion. Which
contribution dominates depends on the propulsion strength, which can be
systematically tuned by changing the concentration of a chemical fuel.  The
mean-square displacements obtained from the experiment exhibit enhanced
diffusion at long lag times.  Our experimental data are consistent with a
Langevin model for the effectively two-dimensional translational motion of an
active Brownian particle in a periodic potential, combining the confining
effects of gravity and the crystalline surface with the free rotational
diffusion of the colloid.  Approximate analytical predictions are made for the
mean-square displacement describing the crossover from free Brownian motion at
short times to active diffusion at long times.  The results are in
semi-quantitative agreement with numerical results of a refined Langevin model
that treats translational and rotational degrees of freedom on the same
footing.
\end{abstract}

\keywords{colloidal microswimmers; active Brownian motion; surface diffusion; hexagonal close-packed monolayer}

\maketitle

\section{Introduction}

The non-equilibrium behaviour of
active and passive particles ranging from microorganisms such as
bacteria and artificial microswimmers to passive colloidal
particles is the focus of a large number of ongoing studies
\cite{lauga2009hydrodynamics, aranson2013active, bechinger2016active, zottl2016emergent}.
Whereas biological microswimmers locomote by means of inherently embedded
nanomotors generating wave-like deformations of their bodies or
appendages, non-biological active particles must be engineered
to support the special conditions to cause self-propulsion.
Passive colloidal particles can be navigated
by external fields or field gradients and can ehxibit nontrivial collective behavior
\cite{loewen2001colloidal,
ghosh2009controlled, dobnikar2013emergent, tierno2014recent,
klapp2016collective}.
In contrast, active colloidal particles, the focus of our study, propel in a
fluid medium also in the absence of the above driving factors; for a review
see, e.g., ref.~\cite{zottl2016emergent}.
By consuming fuel or energy, they typically create local field gradients by themselves,
leading to self-propulsion, while being subjected to rotational Brownian diffusion
\cite{paxton2004catalytic, howse2007self, jiang2010active, buttinoni2012active, kroy2016hot}.

The motion of a particle can be significantly affected by the
presence of a confining boundary.
Due to hydrodynamic coupling, the mobility of a passive particle dragged or
rotating in the vicinity of a plane wall is significantly suppressed
\cite{happel2012low}.
In addition, active motion near surfaces \cite{palacci2013living,
zottl2014hydrodynamics, schwarz2012phase, ma2015catalytic, das2015boundaries,
uspal2015self, simmchen2016topographical, nourhani2016spiral} or other
boundaries such as fluid interfaces \cite{malgaretti2016active}
is complicated by the swimmer--wall interaction depending in general on the
detailed properties of the swimmer, the wall, and the fluid \cite{zottl2016emergent}.
For instance, the concentration of chemical fields near self-phoretic swimmers can be
modified by the presence of a surface. Further,
active particles tend to accumulate at surfaces, even in the
absence of direct, e.g., attractive electrostatic, interactions between the
swimmer and the surface \cite{schaar2015detention}.

Apart from surfaces, colloidal particles have also been confined by imposing
external potentials.
The transport properties of passive colloidal particles have recently been
shown to change when driven over one- \cite{evstigneev2008diffusion,
straube2013synchronous, juniper2016colloidal, pelton2004transport} and
two-dimensional
\cite{pelton2004transport,kim2009activated,ma2013colloidal,ma2015colloidal,su2017colloidal-jcp}
spatially periodic potential landscapes.
Further complexity arises for time-dependent \cite{tierno2012antipersistent,
martinez2016bidirectional, tierno2016enhanced, brazda2017experimental} and
spatially random potentials \cite{evers2013particle, su2017colloidal-sm,
skinner2013localization, schnyder2015rounding, leitmann2017time}.
Depending on the details of driving mechanism, the use of such landscapes can result in
the possibility to precisely control the speed of the net motion
\cite{straube2013synchronous, juniper2016colloidal, martinez2016bidirectional,
brazda2017experimental}, the strength of the diffusion
\cite{evstigneev2008diffusion, ma2013colloidal,ma2015colloidal,
su2017colloidal-sm} and the appearance of transport anomalies
\cite{tierno2012antipersistent, tierno2016enhanced, skinner2013localization,
schnyder2015rounding, leitmann2017time}.
For active colloids confined by external potentials, it has been found that, in
certain cases, they behave similarly to passive particles with an elevated effective
temperature \cite{palacci2010sedimentation, geiseler2016kramers} or subject to an
effective potential \cite{pototsky2012active}.
Initial simulation studies of microswimmers exploring a heterogeneous, random landscape
\cite{schirmacher2015anomalous, zeitz2017active} suggest a rich phenomenology
due to the interference of the landscape with the persistence properties of the trajectories.

In this combined experimental and theoretical study, we investigate the
interplay of active propulsion and a periodic confining potential.
Experimentally, active colloidal micro-spheres \cite{howse2007self,
gao2013organized, choudhury2015surface}
are moving over a periodic surface realized as a hexagonal close-packed (HCP)
monolayer of colloidal particles.
The particles' activity is controlled by changing the concentration of a
chemical propellant.
Theoretically, the three-dimensional motion of an active colloid over the
crystalline surface is treated as active Brownian motion in a two-dimensional
energy landscape, while also accounting for the particles' rotational diffusion.
We demonstrate an intricate interplay between confinement effects and active
motion, leading to non-trivial dependencies of long-time diffusion coefficients
and crossover timescales.

The paper is outlined as follows:
in \cref{sec:experiment}, we describe the experimental system and analysis methods and infer
quantitative estimates of model parameters.
The aspect of active propulsion over a planar surface is discussed in \cref{sec:active-planar}.
In \cref{sec:active-cryst}, we proceed to the general case of propulsion over
the crystalline surface, which includes the derivation of analytic approximations,
the analysis of experimental data, and numerical simulations. We conclude by summarizing
our findings in \cref{sec:conclusion}.

\section{Experimental system}
\label{sec:experiment}

\subsection{Materials and Methods}

\begin{figure}
  \includegraphics[width=75mm]{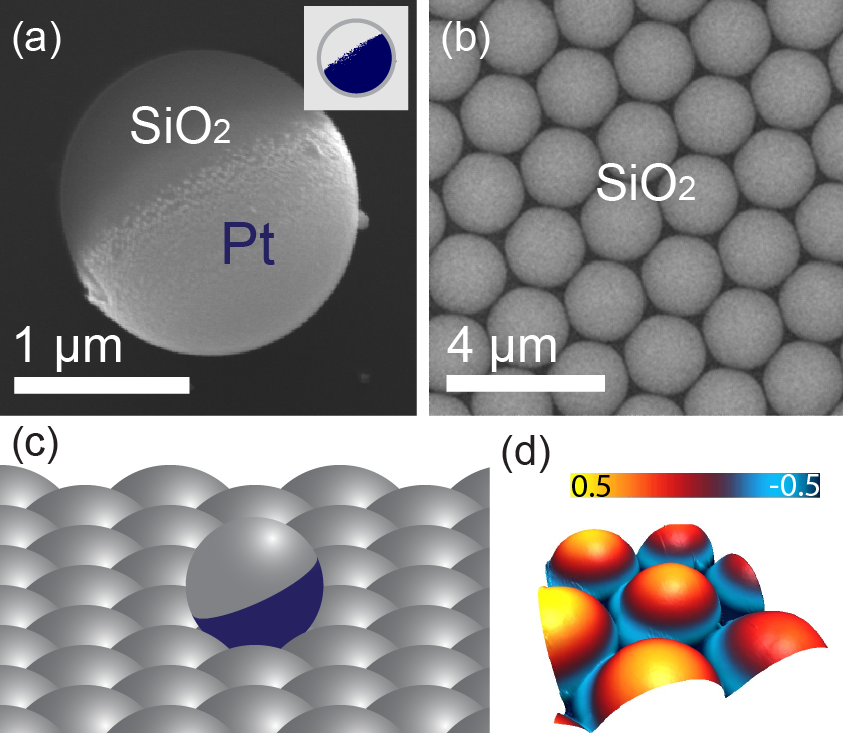}
  \caption{
  (a) Scanning electron microscope (SEM) image of a single half-coated Janus particle;
      inset: the dark-blue shows the location of the Pt cap. \quad
  (b) Top-view SEM image of an HCP monolayer of \ce{Si O_2} microbeads.\quad
  (c) An oblique-view schematic of Janus particle situated on the periodic, two-dimensional lattice,
      giving a sense of the corrugated, periodic morphology of the surface. \quad
  (d) Atomic force microscope (AFM) image exhibiting the topography of the surface,
      color indicates the height in \si{\micro m}.
  }
  \label{scheme}
\end{figure}

First we describe the experimental system and how the activity of the particles
may be tuned by changing the concentration of a chemical propellant, as shown in several previous reports \cite{paxton2004catalytic, howse2007self, gibbs2009design}.
A hexagonal close-packed (HCP) monolayer consisting of
spherical silica (\ce{Si O_2}) microbeads (average diameter $d=\SI{2.07}{\micro m}$ with a coefficient of variation of 10--15\%, Bangs Laboratories) forms the periodic surface upon which the active colloids move;
the lattice constant of the crystal is set by the particle diameter.
The HCP monolayer was prepared with a Langmuir-Blodgett (LB)
deposition technique \cite{bardosova2010langmuir} and covered an entire silicon wafer.
A scanning electron microscope (SEM) image of the monolayer can be seen in \cref{scheme}(b),
and the actual topography of the surface is inferred from the atomic force microscope (AFM) image in \cref{scheme}(d).

The silica microspheres were first functionalized with allyltrimethoxysilane then dispersed in chloroform.
This colloidal suspension was then distributed over the air--water interface of an LB trough.
A cleaned silicon wafer is dipped into the trough and, upon slowly pulling out the wafer,
the monolayer is compressed to form a close-packed assembly.
This process transfers the monolayer from the air--water interface to the silicon wafer.
The wafer is then dried and treated with air plasma to remove any organic impurities
before the experiments.
While the LB technique yields large area HCP monolayers of silica beads,
microscopic line defects can result from the lattice mismatch between adjacent self-assembled
colloidal crystals \cite{szekeres2003two}. In order to ensure consistency of the underlying substrate topography,
the lattice experiments were carried out on the same piece of wafer by varying
the peroxide concentration for the same batch of particles.

The active colloids were fabricated by evaporating a \SI{2}{nm} Cr adhesion layer followed by \SI{5}{nm}
of Pt onto microbeads of the same type as used for the monolayer; see \cref{scheme}(a) for an SEM image.
The thus formed Janus spheres were
then suspended into \ce{H_2 O} and subsequently pipetted onto an HCP lattice surface [\cref{scheme}(c)].
The Pt on the Janus
particle catalyzes the decomposition of hydrogen peroxide (\ce{H_2 O_2}) and thus gives rise to
self-propulsion \cite{howse2007self}. The strength of the propulsion was altered by adding different
concentrations of aqueous \ce{H_2 O_2} to the colloidal suspension, covering concentrations between $0$
and \SI{6}{\%~(v/v)}.
For each concentration, trajectories from 10 randomly chosen Janus particles were recorded for \SI{100}{s}
at a frame rate of \SI{10}{fps} with a Zeiss AxioPhot microscope in reflection mode with a 20$\times$ objective coupled to a CCD camera (pixel size $\SI{5.5}{\micro m} \times \SI{5.5}{\micro m}$, resolution $2048 \times 1088$).

\subsection{Data analysis}
\label{sec:data-analysis}

We computed time-averaged mean-square displacements (MSDs) of 10 trajectories for each \ce{H_2 O_2} concentration, and by averaging the MSDs at each lag time we obtain the ``averaged MSD'' and its standard error. Data fitting was performed with
the software OriginLab (OriginLab Corp., Northampton, MA) using a Levenberg--Marquadt iteration algorithm. Due to the linearly
spaced time grid, the data points accumulate in the double-logarithmic representation at large times.
To account for the different density of data points at short and long lag times on logarithmic scales, we used $1/t$ as weighting factor.
The fits were then performed simultaneously for all 10 data sets of each concentration such that
the different scatter of the data points enters the error estimate of the fit parameters.
The value of the free diffusivity $D_0$ at different \ce{H_2 O_2} concentrations was slightly
adjusted afterwards to obtain the best match with the averaged MSD curves.

\subsection{Height of the potential barrier}

\begin{figure}
  \includegraphics[width=\textwidth]{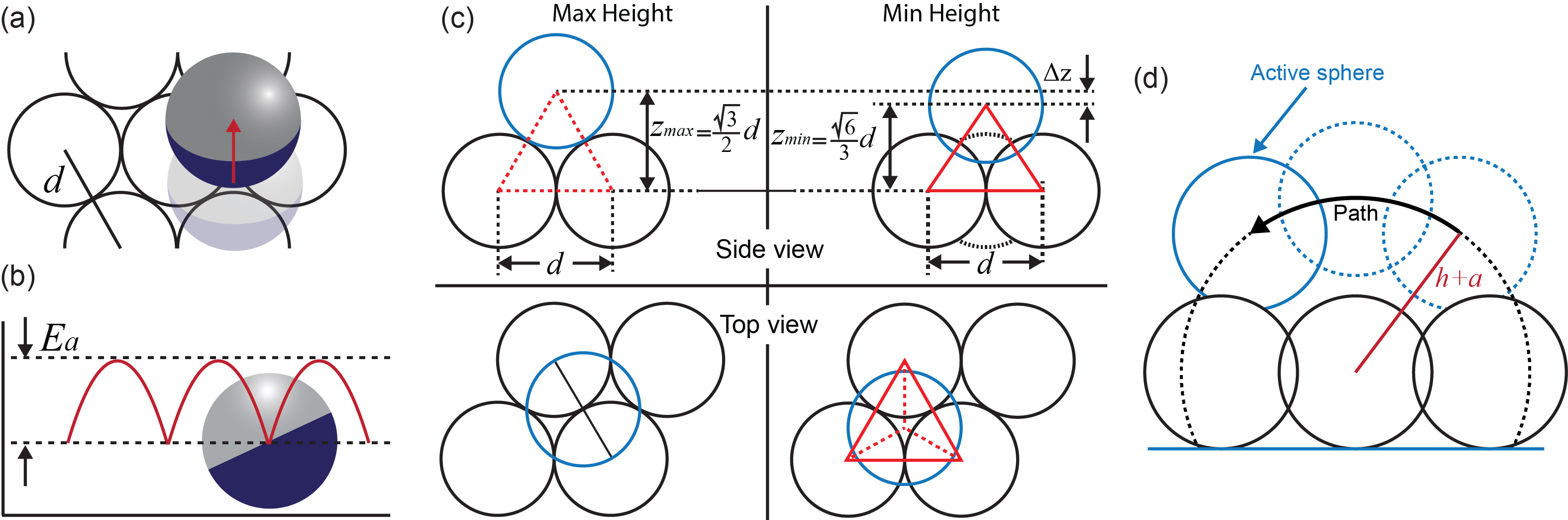}
  \caption{
  (a) Top-view schematic of a single hop from one potential well to an adjacent one. \quad
  (b) Side-view schematic of an active particle situated in an energy minimum. \quad
  (c) Geometry of the barrier between adjacent potential wells due to the HCP lattice substrate.
  Top~row: side-view of the mobile particle (blue circle) at its highest and lowest out-of-plane positions,
  $z_\text{max}$ and $z_\text{min}$ on the left and right columns, respectively.
  Bottom row: corresponding top-views, representing the in-plane positions of the mobile particle
  with respect to the location of the substrate particles (black circles).  \quad
  (d) A cross-sectional representation of the Janus sphere traversing the energy barrier
  as it moves from one adsorption site to an adjacent one. The center of the sphere follows the black curved arrow,
  which is a section of a circle of radius $h+a$, where $h$ is the ``elevation'' above the crystalline surface.
  The maximum position corresponds to the dashed light-blue circle in the top-center portion of the figure.
  }
  \label{max}
\end{figure}

Under gravity the Janus particles settle onto the substrate; once settled, Brownian motion leads to
effectively two-dimensional diffusion in the gravitational potential imposed by the surface.
The potential exhibits a periodic, hexagonal structure of potential wells with adjacent energy minima
separated by a distance $d/\sqrt{3}$. A series of ``hops'' are observed between adjacent energy minima,
or in analogy to surface diffusion, adjacent ``adsorption sites.'' \Cref{max}(a) schematically
demonstrates a single hop from one minimum to an adjacent one.
A successful hop requires the Janus particle to overcome an energy barrier of height $E_\text{a}$,
as depicted in \cref{max}(b).

The gravitational potential $U(\vec x) = \Delta m\, g \, z(\vec x)$ is given by
the buoyant mass $\Delta m$ of the Janus particle, the acceleration $g$ due to gravity,
and the height profile $z(\vec x)$ at the two-dimensional position $\vec x$.
The energy barrier between adjacent potential minima is thus $E_\text{a} = \Delta m \, g \, \Delta z$,
where $\Delta z$ follows from elementary geometry as shown in \cref{max}(c) for
the configurations of maximal and minimal height. At the barrier maximum (left column of \cref{max}(c)),
the centers of two substrate particles and the mobile one form an equilateral triangle.
The Janus particle is thus elevated by $z_\text{max} = \sqrt{3} d / 2$ above the centers
of the substrate particles. If the Janus particle is found in a potential minimum,
the centers of three substrate particles and the mobile one form a regular tetrahedron,
thus $z_\text{min}=\sqrt{6} d / 3$.
For a successful hop, the particles must overcome a geometric barrier of height
$\Delta z=z_\text{max}-z_\text{min} \approx d / 20$,
which evaluates to $\Delta z \approx \SI{100}{nm}$ for $d=\SI{2}{\micro m}$.
This is consistent with the surface's height profile obtained from AFM, see \cref{scheme}(d).

The second ingredient to the energy barrier $E_a$ is the total force on the Janus particle,
which results from the competition of gravitation and buoyancy,
\textit{i.e.}, the energy barrier is also a function of the material from which the Janus particle is made.
Let us first consider a silica sphere which has no metal coating. Then the buoyant mass
is $\Delta m = \Delta \rho V_{\ce{Si O_2}}$, where $\Delta \rho = \rho_{\ce{Si O_2}} - \rho_{\ce{H_2 O}}$
is the difference in density between \ce{Si O_2} and \ce{H_2 O}, and
$V_{\ce{Si O_2}}=(4/3) \pi a^3$ is the volume of the fluid displaced by the particle of radius $a=d/2$.
For a bare \ce{Si O_2} bead of \SI{2}{\micro m} diameter, this yields $E_\text{a} =1.7\,\kB T$
in terms of thermal energy $\kB T$.

In the case of the Janus particle, the asymmetric distribution of the metallic coating needs to be taken into account.
Even though the volume of the cap is small in comparison to that of the bead, it has
a significant effect on $E_a$ due to the higher density of \ce{Pt}.
Following ref.~\citenum{campbell2013gravitaxis},
we model the hemispherical cap as ellipsoidal in shape; the thickness $\Delta a$ of the deposited metal
is largest at the top of the sphere and tapers to zero at the equator. This assumption is justified from the deposition process, which delivers the atoms in the vapor plume ballistically to
the surface of the sphere. The volume of the cap then reads
$V_{\ce{Pt}} = \bigl[ (4/3) \pi a^2(a + \Delta a) - (4/3) \pi a^3] / 2 = (2/3) \pi a^2 \Delta a$,
and with this, the buoyant mass of the Janus particle is
$\Delta m = ({\rho_{\ce{Si O_2}} - \rho_{\ce{H_2 O}}}) V_{\ce{Si O_2}} + ({\rho_{\ce{Pt}} - \rho_{\ce{H_2 O}}}) V_{\ce{Pt}}$.
Adopting a value of $\Delta a =5$\,nm as the maximal thickness of the \ce{Pt} cap and using
$\rho_{\ce{Pt}} =\SI{21.4}{g/cm^3}$, we estimate $E_\text{a} = \Delta m \,g \,\Delta z \approx 2.1\,\kB T$ for the energy barrier of
the Janus particle.

\subsection{Distance of the particle to the surface}
\label{sec:distance}

For the Janus particle moving passively over a smooth plane, we have measured for the translational
diffusion constant $D_0=0.13~\mu$m$^2$/s, which implies a translational
(hereafter indicated by the subscript `$T$') hydrodynamic friction of
$\zeta_\text{T}=\kB T/D_0 = \SI{3.2e-8}{Pa.s.m}$ at $T=\SI{298}{K}$.
As expected, the presence of
a surface increases the hydrodynamic friction compared to unbounded motion: comparing with
the Stokes friction $\zeta_\text{T}^\text{St}=6\pi \eta a \approx \SI{1.74e-8}{Pa.s.m}$ in
\ce{H_2 O} ($\eta=\SI{0.89}{mPa.s}$), we find $\zeta_\text{T} \approx 1.8\,\zeta_\text{T}^\text{St}$.
For a planar surface, the friction coefficient $\zeta_\text{T}$ of a sphere of radius $a$
dragged parallel to the surface at a distance $h$ from the sphere center
obeys Faxén's famous result \cite{goldman1967slow,happel2012low}:
\begin{equation}
    \frac{\zeta_\text{T}^\text{St}}{\zeta_\text{T}} \simeq 1-\frac{9}{16}\frac{a}{h} +\frac{1}{8} \left(\frac{a}{h}\right)^3
    -\frac{45}{256} \left(\frac{a}{h}\right)^4-\frac{1}{16} \left(\frac{a}{h}\right)^5 \,,
    \quad a \ll h,
    \label{faxen}
\end{equation}
Inserting the above experimental value for $\zeta_\text{T}$ and solving for $h$ with $a=\SI{1}{\micro m}$, we obtain $h \approx \SI{1.3}{\micro m}$
leaving a gap of $h - a\approx \SI{0.3}{\micro m}$ between the two surfaces of the Janus particle and the planar substrate.

Note that Faxén's calculation relies on a far-field expansion of the flow field and is justified only for small ratios $a/h$.
As can be seen \textit{a posteriori} we have $a/h \approx 0.8$,
implying slow convergence. Indeed, truncating after the 3\textsuperscript{rd} order in $a/h$ yields an unphysical $h < a$,
which is fixed by the 4\textsuperscript{th} order term. The 5\textsuperscript{th} order term yields merely a relative correction of 2\%, which suggests convergence of the series.
In the following, we anticipate that the translational friction $\zeta_\text{T}$ does not change
appreciably for the range of \ce{H_2 O_2} concentrations used, although it may be modified for active
motion due to altered boundary conditions at the colloid's surface \cite{uspal2015self}.

\section{Active motion on a plane and enhanced diffusion}
\label{sec:active-planar}

In order to control the activity, we exploit the \ce{H_2 O_2}-concentration dependence
of active motion seen in catalytic chemical self-propulsion \cite{howse2007self}.
On a flat, planar surface, the MSD after a lag time $t$ is given by \cite{howse2007self}
\begin{equation}
    \Delta R^2_\text{a,p}(t) = 4\Big( D_0 + \tfrac{1}{2}v^2 \tau_\text{rot} \Big)t +2v^2 \tau_\text{rot}^2\big(\e^{-t/\tau_\text{rot}}-1\big),
    \label{MSD}
\end{equation}
which follows upon assuming independence of translational diffusion of the colloid center and rotational diffusion of the \ce{Pt} cap orientation;
here $D_0$ is the diffusion coefficient for passive Brownian motion over a smooth plane,
$v$ the root-mean-square propulsion velocity in the plane, and $\tau_\text{rot}$ the persistence time of the propulsion direction.
The subscripts `$a$', `$0$' are used to distinguish between the \textit{active} and \textit{passive} motion
and `$p$' refers to the case of \textit{planar} surface.
Qualitatively, \cref{MSD} implies that the active colloid undergoes passive diffusion for $t \ll \tau_{0} := 4D_0/v^2$,
ballistic motion for $\tau_0 \ll t \ll \tau_\text{rot}$, and enhanced diffusion for $t \gg \tau_\text{rot}$.
In the latter regime, the MSD grows linearly with time with an increased diffusion coefficient
$D_\text{a,p}=D_0+v^2 \tau_\text{rot}/2$. Thus at long times, the motion of the active particle displays an enhanced diffusion
relative to the motion of the passive particle ($v=0$).

\begin{figure}
  \includegraphics[width=80mm]{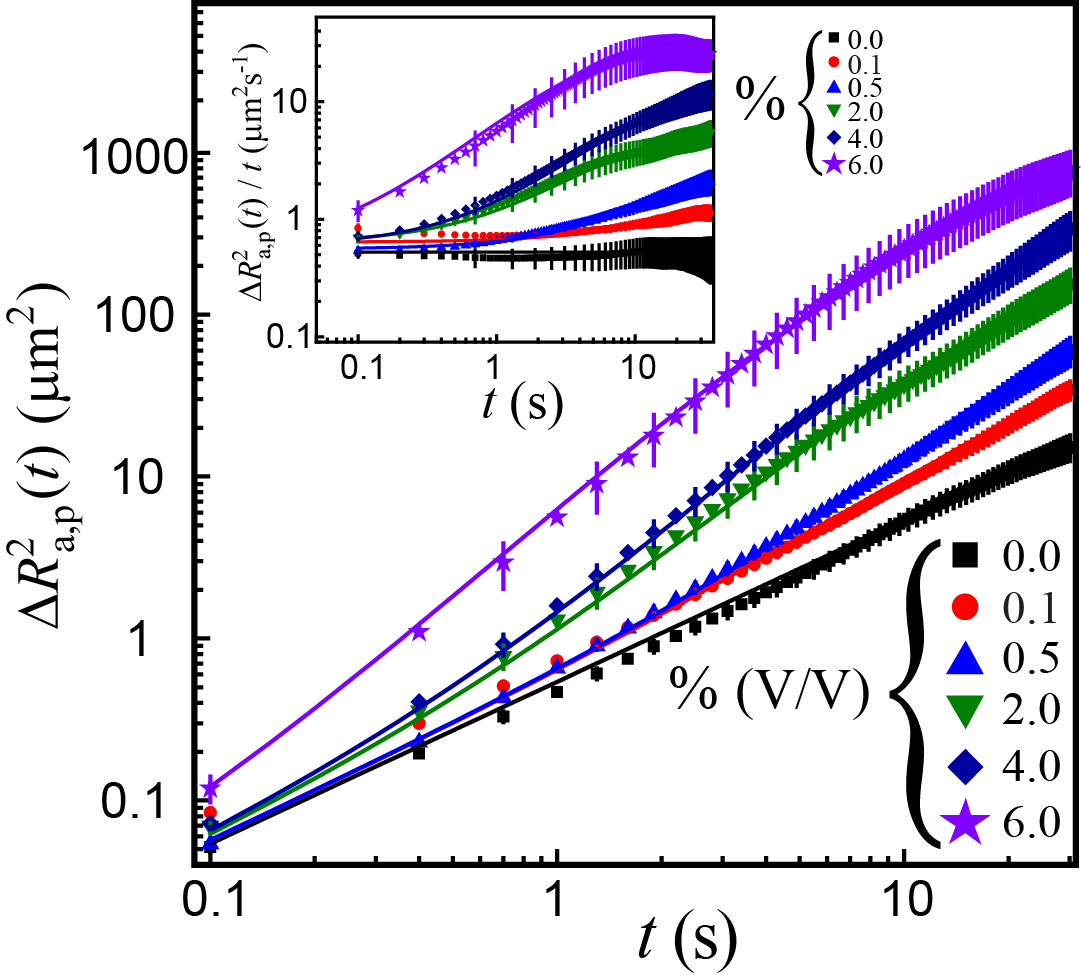}
  \caption{Experimental MSDs of the active particles moving over a planar surface
  for six \ce{H_2 O_2} concentrations (v\%/\%). Solid lines are fits to \cref{MSD}. Inset: rectification plot of the same data
  showing $\Delta R^2_\text{a,p}(t)/t$ vs.\ $t$ in order to more clearly expose the crossover
  from Brownian to enhanced diffusion. It also serves as a test of the fit quality.}
  \label{plane}
\end{figure}

\Cref{plane} shows the experimental results for active motion on a smooth, planar surface
for the six \ce{H_2 O_2} concentrations investigated,
from which the enhanced diffusion was obtained.
The solid curves in \cref{plane} are fits to \cref{MSD}, following the procedure in \cref{sec:data-analysis},
with the obtained parameters given in \cref{table_plane}.
In the inset of \cref{plane}, we have rectified the MSD by dividing by the time lag~$t$. This way, the crossover from
Brownian diffusion $D_0$ at short times to enhanced diffusion $D_\text{a,p}$ at long times can be inferred more easily,
and this representation serves also as a sensitive test of the fit quality.

\begin{table}
  \renewcommand\tabularxcolumn[1]{>{\hfill}p{#1}<{\hfill\hbox{}}}
  \heavyrulewidth=.1em
  \begin{tabularx}{.8\textwidth}{d{5}d{10}d{10}cd{10}}
\toprule
  \multicolumn{1}{X}{$c$ (\si{\% (v/v)})} &
  \multicolumn{1}{X}{$D_0$ (\si{\micro m^2/s})} &
  \multicolumn{1}{X}{$D_\text{a,p}$ (\si{\micro m^2/s})} &
  \multicolumn{1}{X}{$\tau_\text{rot}$ (\si{s})} &
  \multicolumn{1}{X}{$v$ (\si{\micro m/s})} \\
\midrule
  0   & 0.13 & 0.13\textsubscript{$\pm$0.05}  & --   & 0 \\
  0.1 & 0.16 & 0.35\textsubscript{$\pm$0.06}  & 11\textsubscript{$\pm$3}    & 0.18\textsubscript{$\pm$0.01} \\
  0.5 & 0.14 & 0.7\textsubscript{$\pm$0.2}    & 12\textsubscript{$\pm$3}    & 0.30\textsubscript{$\pm$0.01} \\
  2   & 0.16 & 1.4\textsubscript{$\pm$0.5}    & 4.2\textsubscript{$\pm$0.7} & 0.77\textsubscript{$\pm$0.04}\\
  4   & 0.15 & 3.6\textsubscript{$\pm$0.8}    & 7.9\textsubscript{$\pm$1.2} & 0.93\textsubscript{$\pm$0.04}\\
  6   & 0.14 & 8.0\textsubscript{$\pm$1.6}    & 2.3\textsubscript{$\pm$0.6} & 2.6\textsubscript{$\pm$0.3}\\
\bottomrule
  \end{tabularx}
  \caption{Parameters obtained from fitting \cref{MSD} to the MSD data for active motion atop a planar substrate, shown in \cref{plane}.
  The long-time diffusion coefficient $D_\text{a,p}$ was calculated from $D_\text{a,p}=D_0+v^2 \tau_\text{rot}/2$.
  The uncertainties are standard errors of the mean obtained from the fitting procedure.}
  \label{table_plane}
\end{table}

The short-time diffusivity $D_0$ was varied between $\SI{0.14}{\micro m^2/s}$ and $\SI{0.18}{\micro m^2/s}$ to obtain
the best match with the averaged MSD curves.
The small variability of the background diffusivity may be attributed
to the sparseness of data points at short timescales, but it may also reflect possibly altered boundary conditions
at the surface due to the catalytic reaction.
We have found that the propulsion velocity $v$ measured this way (\cref{table_plane}) increases monotonically with the \ce{H_2 O_2} concentration~$c$.
Similarly, the long-time diffusivity $D_\text{a,p}$ grows progressively with the increase in $c$ and is enhanced over $D_0$ for all concentrations $c > 0$ studied, in accordance with \cref{MSD}.
We remark that $v$ and $D_\text{a,p}$ for $c=\SI{6}{\%~(v/v)}$ are significantly larger than the corresponding values for $c=\SI{4}{\%~(v/v)}$,
suggesting that additional effects become important for the propulsion mechanism at this high concentration.
Finally, we observe a large, non-monotonic variation of $\tau_\text{rot}$, signifying that
the rotational motion is non-trivially altered by the activity.
We attribute this to imperfections in the Janus particle, causing deviations
from axisymmetric symmetry and thus the possibility of a residual active
angular velocity on the particle.

\section{Active motion atop a crystalline surface}
\label{sec:active-cryst}

\subsection{Theory}

The behavior of the active colloids moving across the crystalline surface
is significantly different from the planar case. The catalyzed chemical reaction on the \ce{Pt} side
leads to two effects described first qualitatively:
\emph{(i)}~similar to the planar case, the Janus particles are actively and directionally propelled
over the surface away from the catalyst coating \cite{he2007designing}.
\emph{(ii)}~In contrast to the planar case, motion over the crystalline surface is hindered by
the particle becoming transiently trapped within potential wells.

According to the above reasoning, the active Brownian motion in the periodic potential $U(\vec{x})$
may be modelled in a simplified way by an over-damped Langevin equation:
\begin{equation}
    \dot{\vec{x}}(t)=\vec{v}(t) - \zeta_\text{T}^{-1} \nabla U\boldsymbol(\vec x(t)\boldsymbol) + \sqrt{2 D_0} \, \vec{\xi}_\text{T}(t).
    \label{langevin}
\end{equation}
Here, $\vec{x}(t)$ and $\vec v(t)$ are, respectively, the vectors of the particle position and of the propulsion velocity
projected onto the plane of the crystalline surface. The latter is incorporated via the second term
on the right hand side of \cref{langevin}.
Further, $\vec{\xi}_\text{T}(t)$ is a two-dimensional Gaussian white noise of zero mean and unit (co-)variance,
$\langle \vec{\xi}_\text{T} (t)  \otimes \vec{\xi}_\text{T} (s)\rangle = \mathbf{I} \, \delta(t-s)$
with $\mathbf{I}$ the identity tensor, to describe passive Brownian motion over a planar surface with
diffusion coefficient $D_0=\kB T/\zeta_\text{T}$.

The partial case of passive Brownian diffusion taking place in a periodic surface potential
is described by \cref{langevin} with vanishing self-propulsion, $\vec{v}=0$:
\begin{equation}
   \dot{\vec{x}}_\text{c}(t) =-\zeta_\text{T}^{-1} \nabla U\boldsymbol(\vec{x}_\text{c}(t)\boldsymbol)+\sqrt{2 D_0} \, \vec{\xi}_\text{T}(t),
    \label{velocity}
\end{equation}
where the subscript `$c$' stands for \textit{crystalline} surface.
The corresponding MSD, $\Delta R_\text{0,c}^2(t) = \mleft\langle |\vec x_\text{c}(t) - \vec x_\text{c}(0)|^2 \mright\rangle$,
is well approximated by a simple exponential memory \cite{fulde1975periodic, risken1978correlation},
which manifests itself in the velocity autocorrelation function (VACF) as
\begin{equation}
    Z_\text{0,c} (t) := \frac{1}{4} \frac{d^2}{dt^2} \Delta R_\text{0,c}^2 (t)
      = D_0 \, \delta(t-0^+)-\frac{\Delta D_\text{c}}{\tau_\text{c}}  \e^{-t/\tau_\text{c}}.
    \label{Z0c}
\end{equation}
Here, $\tau_\text{c}$ is the longest relaxation time of the process and $\Delta D_\text{c} > 0$ describes the reduction of the long-time
diffusivity due to the presence of the periodic surface relative to the planar case.
This ansatz for the VACF corresponds to keeping only the largest non-zero eigenvalue of the Smoluchowski operator~\cite{festa1978diffusion}.
For the MSD, one readily calculates:
\begin{align}
    \Delta R_\text{0,c}^2(t) = 4 \int_{0}^{t}(t-s) \, Z_\text{0,c}(s) \, \diff s
    = 4(D_0-\Delta D_\text{c})t -4\Delta D_\text{c} \tau_\text{c} (\e^{-t/\tau_\text{c}}-1) \,.
    \label{msd-0c}
\end{align}
It describes a simple crossover from free, unconfined diffusion with $D_0=\kB T/\zeta_\text{T}$ at short times
($t \ll \tau_\text{c}$) to diffusion at long times with a reduced diffusion constant
$D_\text{0,c} = D_0-\Delta D_\text{c} \le D_0$ for $t \gg \tau_\text{c}$.
The crossover timescale $\tau_\text{c}$ describes the time after which the particle has explored
a single potential minimum. Its inverse, $\tau_\text{c}^{-1}$, may be interpreted as
the attempt rate for escaping from the potential well \cite{kramers1940brownian, festa1978diffusion}.

Next, we combine this result for passive motion in a periodic potential with active motion
under the assumption that the diffusive motion in the potential be independent of the direction of the propulsion velocity.
More precisely, we require that $\vec{x}_\text{c}(t)$ and $\vec{v}(t)$ are decorrelated on the scale
of $\tau_\text{c}$, noting that merely the increment $\vec{\xi}_\text{T}(t)$ is strictly independent of $\vec{v}(t)$.
Then, the autocorrelation of the propulsion velocity is simply added to the VACF for passive diffusion,
\begin{equation}
    Z_\text{a,c}(t)=Z_\text{0,c}(t)+\tfrac{1}{2} \langle \vec{v}(t)\cdot \vec{v}(0) \rangle \,.
    \label{Zac}
\end{equation}
Rotational diffusion of the cap orientation determines the
propulsion velocity vector $\vec{v}(t)$ projected onto the surface plane.
Neglecting the small gravitational torque on the present Janus particles (see also \cref{sec:langevin}), it follows that
\begin{equation}
\langle \vec{v}(t)\rangle=0, \quad \langle \vec{v}(t)\cdot \vec{v}(0)\rangle =v^2 \e^{-t/\tau_\text{rot}} \,,
    \label{vel-corr}
\end{equation}
with $\tau_\text{rot}$ the persistence time of the orientation;
for free three-dimensional rotation $\tau_\text{rot} = \mleft(2 D_\text{rot}\mright)^{-1}$.
With this, the MSD of an active particle moving atop a crystalline surface follows from \cref{Zac,Z0c}, again by integration:
\begin{equation}
  \Delta R_\text{a,c}^2(t)
    = 4\mleft( D_0-\Delta D_\text{c}+\tfrac{1}{2}v^2\tau_\text{rot} \mright) t
    - 4\Delta D_\text{c} \, \tau_\text{c} \mleft( \e^{-t/\tau_\text{c}}-1 \mright) \\
         + 2 v^2 \tau_\text{rot}^2 \mleft( \e^{-t/\tau_\text{rot}}-1 \mright).
  \label{MSD_full}
\end{equation}
In \cref{sec:langevin}, this prediction will be checked against simulations.

\subsection{Experiment}

\begin{figure}
  \includegraphics[width=80mm]{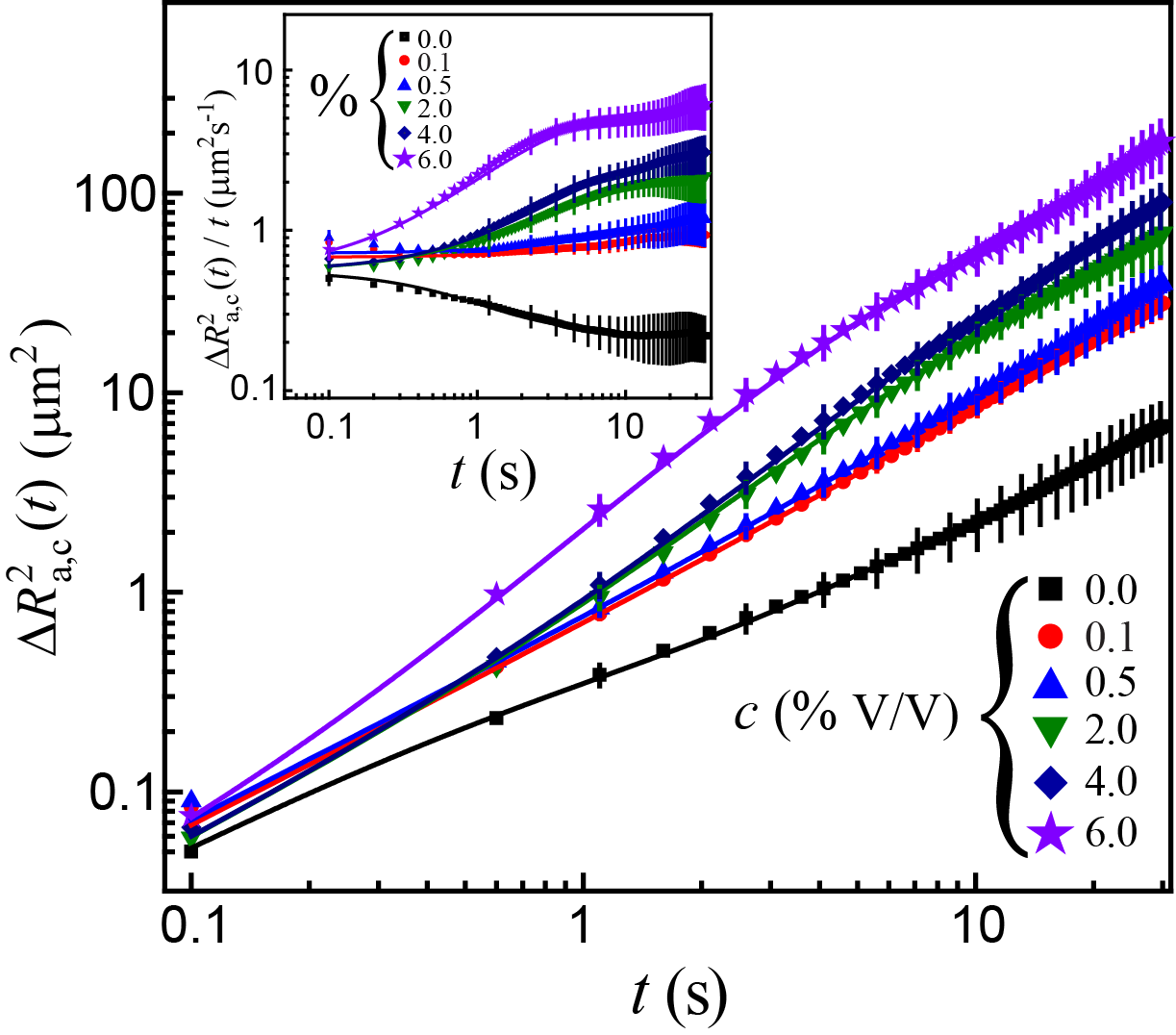}
  \caption{MSDs of the active particles moving atop the crystalline surface for six \ce{H_2 O_2} concentrations \%~(v/v).
  The solid lines are fits to \cref{MSD}. Inset: rectification of the same data by plotting $\Delta R_\text{a,c}^2(t) / t$ vs. $t$}
  \label{MSD_lattice}
\end{figure}

\Cref{MSD_lattice} shows the averaged MSD data of Janus particles being actively propelled atop the crystalline surface for six \ce{H_2 O_2} concentrations.
As expected, increasing the \ce{H_2 O_2} concentration leads to higher observed propulsion speeds
and higher long-time diffusion.
The latter can be directly inferred from the rectification $\Delta R_\text{a,c}^2(t) / t$ displayed in the inset of \cref{MSD_lattice}.
The data suggest further a monotonic dependence on time, either decreasing or increasing depending on the concentration of fuel,
which we interpret as a competition of the suppression of diffusivity due to the potential landscape with the enhancement due to active motion.

Fitting \cref{MSD_full} to the experimental MSD data would, in principle, provide an estimate for the parameters
$\Delta D_\text{c}$, $v$, $\tau_\text{c}$, $\tau_\text{rot}$. Following this approach,
it turned out all of the parameters depend on the \ce{H_2 O_2} concentration.
Specifically, fixing the values of $v$ and $\tau_\text{rot}$ to those from the experiments
with the planar surface does not produce satisfying fits.
Further, the four-parameter fits suggest similar
values for $\tau_\text{c}$ and $\tau_\text{rot}$, which motivated us to merge
both timescales into a single parameter, $\tau$.
This is consistent with the absence of any minimum or maximum at intermediate
lag times in the data for $\Delta R_\text{a,c}^2(t) / t$, which would be supported by \cref{MSD_full}.
However, $\tau_\text{c} \approx \tau_\text{rot}$ implies that the parameters $\Delta D_\text{c}$ and $v$ are no longer independent,
merely the combination $\Delta D=\Delta D_\text{c}-v^2 \tau /2$ can be obtained.
Thus, \cref{MSD_full} reduces to a simplistic, effective model of
the MSD of a self-propelled particle atop a periodic surface,
\begin{equation}
    \Delta R^2_\text{a,c} (t) \approx 4D_\text{a,c} t+4(D_\text{a,c}-D_0)\tau  \, \mleft( \e^{-t/\tau}-1 \mright).
    \label{lat}
\end{equation}
By construction, this result has the same form as \cref{MSD,msd-0c}, but the interpretation of the parameters is different in each case.
For long times ($t \gg \tau_\text{c},\tau_\text{rot}$), the MSD increases linearly,
and the combination $D_\text{a,c} := D_0-\Delta D_\text{c}+v^2 \tau/2$
is the long-time diffusion coefficient on the crystalline surface.

\begin{table}
  \renewcommand\tabularxcolumn[1]{>{\hfill}p{#1}<{\hfill\hbox{}}}
  \heavyrulewidth=.1em
  \begin{tabularx}{.6\textwidth}{d{5}d{8}d{8}d{8}d{6}}
\toprule
  \multicolumn{1}{X}{$c$ (\si{\% v/v})} &
  \multicolumn{1}{X}{$D_0$ (\si{\micro m^2/s})} &
  \multicolumn{1}{X}{$D_\text{a,c}$ (\si{\micro m^2/s})} &
  \multicolumn{1}{X}{$\tau$ (\si{s})} \\
\midrule
  0   & 0.14\textsubscript{$\pm$0.01} & 0.05\textsubscript{$\pm$0.01}  & 0.4\textsubscript{$\pm$0.2}  \\
  0.1 & 0.17\textsubscript{$\pm$0.03} & 0.25\textsubscript{$\pm$0.03}  & 6.3\textsubscript{$\pm$2.9}  \\
  0.5 & 0.18\textsubscript{$\pm$0.02} & 0.29\textsubscript{$\pm$0.02}  & 5.4\textsubscript{$\pm$2.2}    \\
  2   & 0.14\textsubscript{$\pm$0.02} & 0.56\textsubscript{$\pm$0.02}  & 2.3\textsubscript{$\pm$0.4}   \\
  4   & 0.14\textsubscript{$\pm$0.03} & 0.8\textsubscript{$\pm$0.5}  & 3.2\textsubscript{$\pm$0.1}   \\
  6   & 0.14\textsubscript{$\pm$0.04} & 1.5\textsubscript{$\pm$0.4}  & 1.5\textsubscript{$\pm$0.4}   \\
\bottomrule
  \end{tabularx}
  \caption{Parameter estimates from fitting \cref{lat} to the MSD data shown in
  \cref{MSD_lattice}. $D_0$ is the Brownian diffusion coefficient at short times,
  $D_\text{a,c}$ is the long-time diffusion coefficient of the self-propelled particle moving atop the crystalling surface,
  and $\tau \approx \tau_\text{c} \approx \tau_\text{rot}$ is a single crossover timescale.
  }
  \label{table_lattice}
\end{table}

We used \cref{lat} to fit the MSD data with $D_0$, $D_\text{a,c}$, and $\tau$ as free parameters.
The results obtained for each concentration are given in \cref{table_lattice} and the fits shown as solid curves
in \cref{MSD_lattice} provide a consistent description of the data. The enhancement of the long-time diffusivity
$D_\text{a,c}$ with increasing \ce{H_2 O_2} concentration is much less pronounced
compared to the case of a planar surface ($D_\text{a,p}$ in \cref{table_plane}),
which is a direct consequence of the trapping potential. We again observe a slight variability
of $D_0$ and a strong dependence of $\tau$ on the \ce{H_2 O_2} concentration.
We note that the MSD for $c=0.1\%$ seems to deviate from the overall trend, and we exclude
this data set from the remaining discussion.

\subsection{Simulation of Langevin equations}\label{sec:langevin}

As an independent check of the above approximate predictions and to gain further insight,
we finally proceed to a refined theoretical model that explicitly includes both the translational and rotational degrees of freedom.
The over-damped dynamics of an active bottom-heavy microswimmer \cite{wolff2013sedimentation} sedimenting due to gravity onto
an HCP monolayer can effectively be described by Langevin equations for the projected position $\vec{x}=(x,y)$
and the orientation $\vec{u}=(u_x,u_y,u_z)$ \cite{oksendal2010stochastic,leitmann2017dynamically}:
\begin{align}
\dot{\vec{x}}(t) & = v_0 \vec{u}_{\parallel} -\zeta_\text{T}^{-1} \nabla U (\vec{x}(t))  + \sqrt{2D_\text{T}} \, \vec{\xi}_\text{T}(t)\, , \label{LE-r}\\
\dot{\vec{u}}(t) & = \left[ \zeta_\text{R}^{-1} \vec{T} + \sqrt{2D_\text{R}} \, \vec{\xi}_\text{R}(t)\right] \times \vec{u} -2D_\text{R} \vec{u}\, , \label{LE-u}
\end{align}
where It\=o's interpretation of the white noise is adopted for the second equation.
The first equation describes translational motion parallel to the surface and simply reproduces \cref{langevin}, where we specify
the propulsion term as $\vec{v} = v_0 \vec{u}_{\parallel}$. Here, $v_0$ is the propulsion strength
and $\vec{u}_{\parallel} := (u_x,u_y)$ is the orthogonal projection of the three-dimensional
unit vector $\vec{u}$ onto the $xy$-plane.
The second equation governs the cap orientation $\vec{u}$ of the Janus particle, in which
$\vec{T} = m r_0 \vec{u} \times \vec{g}$ is the gravitational torque
with $m=V_{\ce{Si O_2}}\rho_{\ce{Si O_2}}+V_{\ce{Pt}}\rho_{\ce{Pt}}$ the mass of the particle and $r_0$
the displacement of the center of mass from the center of the sphere due to the heavy cap;
for the Janus particle used in the present experiments, $r_0 \approx 0.02 \,a$.
The strength of thermal fluctuations is determined by $D_\text{T} = \kB T/\zeta_\text{T}$
and $D_\text{R} = \kB T /\zeta_\text{R}$,
the diffusivities of the translational and rotational motions, respectively, with $\vec{\xi}_\text{T}$
and $\vec{\xi}_\text{R}$ being independent Gaussian white noises having zero mean and unit covariance.
The effective substrate potential $U(x,y)=\Delta m\, g \, z(x,y)$ arises from
the buoyancy-corrected graviational force $\Delta m\, g$ on the Janus particle with its height given by the landscape $z(x,y)$.
In this effectively two-dimensional representation, the center of the Janus particle
is constrained to the surface $z=z(x,y)$, which is created by the HCP colloidal monolayer
and is composed of the upper non-intersecting segments of spheres,
\begin{align}
  z(x,y) = \sqrt{(h+a)^2 - (x-x_i)^2 - (y-y_i)^2},
\end{align}
located at the centers $(x_i, y_i)$ of the monolayer particles, which form
a hexagonal lattice of lattice constant $d=2a=\SI{2}{\micro m}$, see \cref{max}(a).
For the sphere radius, we use $h + a \approx \SI{2.3}{\micro m}$ as estimated in \cref{sec:distance}, see also \cref{max}(d).

The presence of the surface also modifies the rotational hydrodynamic friction. With the above value of the elevation $h$,
we estimate for the rotational diffusion constants, which are distinct for rotations parallel and perpendicular
to the surface \cite{happel2012low,leach2009comparison}:
\begin{equation}
  \frac{D_\text{R}^\parallel}{D_\text{R}^\text{St}} \approx 1-\frac{1}{8}\left(\frac{a}{h}\right)^3, \qquad
  \frac{D_\text{R}^\perp}{D_\text{R}^\text{St}} \approx 1-\frac{5}{16}\left(\frac{a}{h}\right)^3,
\end{equation}
where $D_\text{R}^\text{St} = \kB T / 8 \eta \pi a^3$. For the sake of simplicity and since we are only interested in the
correlation $\langle \vec u(t)\cdot \vec u(0) \rangle$, we further use the average
rotational diffusion constant
\begin{equation}
  D_\text{R} := \frac{1}{3} \mleft(D_\text{R}^\parallel + 2 D_\text{R}^\perp \mright)
    = \mleft[ 1 - \frac{1}{4} \mleft(\frac{a}{h}\mright)^3 \mright] D_\text{R}^\text{St}\,,
\end{equation}
such that $D_\text{R} \approx \SI{0.15}{s^{-1}}$ and $\tau_\text{rot} = (2 D_\text{R})^{-1}\approx \SI{3.3}{s}$.
With this, the model \cref{LE-r,LE-u} are fully specified with $v_0$ as a control parameter and
is solved numerically with the standard Euler-Maruyama scheme.
We have validated the numerical scheme for the planar case ($U=0$) by comparing simulated MSDs with the exact solution, \cref{MSD}, for this case.

First of all we note that \cref{LE-u} for the orientational dynamics is decoupled from the translational motion, \cref{LE-r}.
Yet, it remains analytically challenging due to the gravitational torque term, $\vec{T}\ne 0$.
The numerical results suggest that the dynamics of the orientation is well approximated by the free solution
$\langle \vec{u}(t)\rangle=0$ and
$\langle \vec{u}(t)\cdot \vec{u}(0)\rangle =\e^{-2 D_\text{R} t}$ and therefore $\tau_\text{rot} = (2 D_\text{R})^{-1}$.
Thus, the problem described by \cref{LE-r,LE-u}
becomes equivalent to the model defined by \cref{langevin} with the velocity correlation prescribed via \cref{vel-corr}.
Note that although the orientation of the particle is three-dimensional,
the translational motion of the particle is essentially restricted to the horizontal plane.
As a result, for the mean-square velocity entering \cref{vel-corr} we have
$v^2 := \mleft\langle ( v_0 \vec u_{\parallel})^2 \mright\rangle = 2 v_0^2/3.$
The latter step tacitly assumes that all orientations of $\vec{v}$ are equally probable,
which is fulfilled under the approximation of free rotational diffusion.

\begin{figure}
\includegraphics[width=0.6\textwidth]{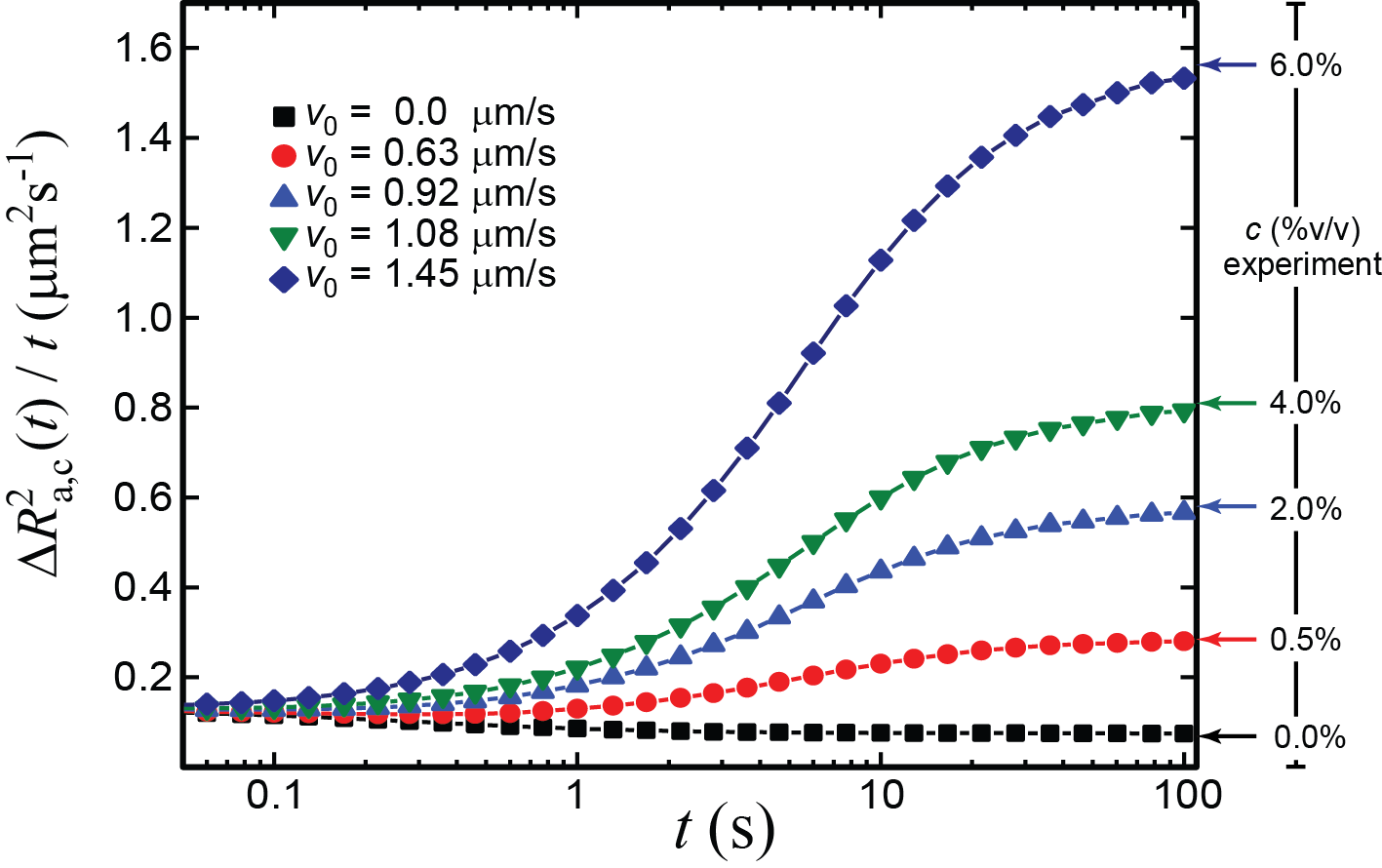}
\caption{The MSDs obtained from simulations (markers) for different values of
$v_0$ for $D_\text{T}=\SI{0.13}{\micro m^2/s}$, $D_\text{R}=\SI{0.15}{s^{-1}}$ and the corresponding fits (solid lines) of \cref{lat}; see also \cref{table_sim}.
In the simulations, $v_0$ is adjusted such that
the experimental long-time diffusion coefficients $D_\text{a,c}$ (arrows) for the different \ce{H_2 O_2} concentrations are approximately recovered.
}
\label{fig-dR2-gen-sim}
\end{figure}

Simulations of the active motion of the Janus particle in the hexagonal landscape,
\cref{LE-r,LE-u}, were performed for different values of $v_0$ with all other parameters
fixed to their values as mentioned above.
The obtained MSDs shown in \cref{fig-dR2-gen-sim} capture the trends
of the experimental data (\cref{MSD_lattice}) semi-quantitatively
and reproduce the long-time diffusion coefficients for propulsion velocities $v=\sqrt{2/3} v_0$ similar to the experimental values.

\begin{table}
  \renewcommand\tabularxcolumn[1]{>{\hfill}p{#1}<{\hfill\hbox{}}}
  \heavyrulewidth=.1em
  \begin{tabularx}{.6\textwidth}{d{5}d{8}d{8}d{8}d{6}}
\toprule
  \multicolumn{1}{X}{$v_0$ (\si{\micro m/s})} &
  \multicolumn{1}{X}{$v$ (\si{\micro m/s})} &
  \multicolumn{1}{X}{$D_\text{a,c}$ (\si{\micro m^2/s})} &
  \multicolumn{1}{X}{$\tau$ (\si{s})} \\
\midrule
  0     & 0     & 0.07  & 0.2  \\
  0.63  & 0.51  & 0.29  & 5.2  \\
  0.92  & 0.75  & 0.58  & 3.2  \\
  1.08  & 0.88  & 0.81  & 3.2  \\
  1.45  & 1.18  & 1.57  & 3.2  \\
\bottomrule
  \end{tabularx}
  \caption{Parameters $D_\text{a,c}$ and $\tau$ obtained from fitting \cref{lat} to the simulated
  MSDs (\cref{fig-dR2-gen-sim}) for different propulsion velocities $v_0$,
  keeping $D_\text{T} = D_0 = \SI{0.13}{\micro m^2/s}$ and $D_\text{R}=\SI{0.15}{s^{-1}}$ fixed.
  For comparison with experiment, the projected propulsion velocity $v=\sqrt{2/3} v_0$ is quoted as well.}
  \label{table_sim}
\end{table}

Finally, the numerical solutions permit a number of insights into the analytic predictions for the MSD.
First, simulations addressing the passive motion above the crystalline surface show that
\cref{msd-0c} is a very good approximation for the experimental regime investigated here. With
$D_\text{T}=\SI{0.13}{\micro m^2/s}$ and $D_\text{R}=\SI{0.15}{s^{-1}}$, excellent quantitative agreement between the
simulated MSD and the theory is found for the parameters $\tau_\text{c}\approx \SI{0.18}{s}$ and
$\Delta D_\text{c}\approx \SI{0.055}{\micro m^2/s}$.
Second, simulations of active motion above the crystalline surface
indicate that the approximation \cref{MSD_full} does not work universally.
For the considered range of propulsion velocities $v_0$, the simulated MSDs are significantly overestimated
by \cref{MSD_full} if $\Delta D_\text{c}$ and $\tau_\text{c}$ are fixed to their values for passive motion ($v_0=0$).
For the cases where $\Delta R_\text{a,c}(t)/t$ is monotonic in $t$, it is, however, possible to obtain good descriptions
if we allow $\Delta D_\text{c}$ and $\tau_\text{c}$ to depend
on the activity, $v_0$, and estimate them by the fit for each value of $v_0$.
Third, for these monotonic cases we have found that \cref{lat} is a good approximation to the MSD,
see \cref{fig-dR2-gen-sim} and \cref{table_sim}. On one hand, it provides less flexibility because
it contains fewer parameters, in particular it allows for only a single crossover. On the other hand,
this is sufficient to capture the full time dependence in these cases
and the formula is simpler to handle than \cref{MSD_full}.
Note that for high \ce{H_2 O_2} concentrations, when $D_\text{a,c}$ is larger than $D_0$,
the crossover time saturates, $\tau \approx \tau_\text{rot}$. For low \ce{H_2 O_2} concentrations, when $D_\text{a,c} \simeq D_0$, $\tau$ may differ from $\tau_\text{R}$ quite significantly.

\section{Conclusions}
\label{sec:conclusion}

We have studied the problem of propulsion of an active colloidal particle above a crystalline surface
by a combination of experiment, theory, and numerical simulations.
The experimental system consists of catalytically driven colloidal Janus spheres sedimenting due to gravity
on top of a periodic substrate. The strength of self-propulsion is controlled by changing
the concentration of the chemical fuel, \ce{H_2 O_2}.
Due to a relatively heavy cap, the center of mass of
the nearly spherical active particle is slightly displaced from its center, which makes it bottom heavy.
The substrate is realized by a HCP colloidal monolayer made of passive stationary colloidal particles of similar size and material.
We have investigated the mean-square displacement of the Janus particle and extracted the parameters characterizing
different regimes of motion. In particular, we looked at the long-time diffusion coefficient, how it changes
relative to the free diffusivity and how it develops from the Brownian motion at short timescales.

We have considered two limiting cases,
which permit comparably simple interpretations, and finally studied their interplay.
First, we have focused on the active propulsion above a planar surface,
which shows an enhanced long-time diffusion constant and is
in agreement with previously known results. Fitting the full time dependence of the MSDs
provides additional details on how activity modifies the rotational diffusion.
Second, we have investigated the case of passive
diffusion above a crystalline surface, which plays the role of a trapping potential and
results in the suppression of the particle diffusivity.
Third, we have studied the interplay of these two factors, which have opposite effects
on the diffusion constant. We show that depending on
the strength of the activity relative to the strength of the trapping potential,
the long-time diffusion constant can be either lower (weak activity) or higher
(strong activity) relative to the free diffusion. In all instances studied, the diffusion
constant of an active particle remains larger than that of a passive particle.

The analytical theory is based on a simplified overdamped Langevin equation, \cref{langevin}.
Based on the fact that the gravitational torque of the particle is weak, we reduce the problem
to the case of a particle whose orientation is subject to free diffusion and
suggest a theoretical formula that describes the mean-square displacement and involves two
generally distinct timescales to describe the twofold crossover from free to active diffusion
due to \emph{(i)} the periodic trapping potential and \emph{(ii)} rotational diffusion.
Further, we show that the present experimental data display a simple crossover
involving only one timescale. Numerical simulations
of the full Langevin model, \cref{LE-r,LE-u}, explicitly addressing both the translational and orientational
degrees of freedom, confirm the expressions for the MSD in the partial cases of active motion above a planar
surface [\cref{MSD}] and for the case of passive motion above the crystalline surface [\cref{msd-0c}]. For the general case,
the suggested analytical formula, \cref{MSD_full}, is shown to serve as an approximation
with only semi-quantitative agreement with the numerical model.

Finally, we note that our analysis of the experiment does not depend
on details of the propulsion mechanism and on the particular realization of the confining potential.
Therefore, we expect that our findings apply equally to a broad class of microswimmers moving in a periodic landscape.




\section*{Acknowledgements}

This work was supported in part by the Deutsche Forschungsgemeinschaft (DFG) as part of
the project SPP 1726 (microswimmers, FI 1966/1-1) and in part by the State of Arizona
Technology and Research Initiative Fund (TRIF), administered by the Arizona Board of Regents.

\bibliography{refs}

\end{document}